\documentclass[aps,prc,preprint,groupedaddress,showpacs]{revtex4-1}
\usepackage{amssymb}
\usepackage{amsmath,bm}
\usepackage[normalem]{ulem}
\usepackage{booktabs}
\usepackage{graphicx}
\usepackage{float}
\usepackage{multirow}
\usepackage{lineno}

\usepackage{tensor}
\usepackage[utf8]{inputenc}
\usepackage{subfigure}

\begin{document}

\title{Enhanced production of strange baryons in high energy nuclear collisions from a multiphase transport model}
\author{Tianhao Shao$^{a,b,c}$}
\author{Jinhui Chen$^{a}$}
\author{Che Ming Ko$^{d}$}
\author{Zi-Wei Lin$^{e}$}

\affiliation{$^a$Key Laboratory of Nuclear Physics and Ion-beam Application (MOE), Institute of Modern Physics, Fudan University, Shanghai 200433, China}
\affiliation{$^b$Shanghai Institute of Applied Physics, Chinese Academy of Science, Shanghai 201800, China}
\affiliation{$^c$University of Chinese Academy of Science, Beijing 100049, China}
\affiliation{$^d$Cyclotron Institute and Department of Physics and Astronomy, Texas A\&M University, College Station, Texas 77843, USA}
\affiliation{$^e$Department of Physics, East Carolina University, Greenville, North Carolina 27858, USA}

\date{\today}
\begin{abstract}

We introduce additional coalescence factors for the production of strange baryons in a multiphase transport (AMPT) model in order to describe the enhanced production of multistrange hadrons observed in Pb-Pb collisions at $\rm \sqrt{s_{NN}}$ = 2.76 TeV at the Large hadron Collider (LHC) and Au+Au collisions at $\rm \sqrt{s_{NN}}$ = 200 GeV at Relativistic Heavy-Ion Collider (RHIC).This extended AMPT model is found to also give a reasonable description of the multiplicity dependence of the strangeness enhancement observed in high multiplicity events in $pp$ collisions at $\rm \sqrt{s}$ = 7 TeV and $p$-Pb collisions at $\rm \sqrt{s_{NN}}$ = 5.02 TeV. We find that the coalescence factors depend on the system size but not much on whether the system is produced from A+A or p+A collisions.  The extended AMPT model thus provides a convenient way to model the mechanism underlying the observed strangeness enhancement in collisions of both small and large systems at RHIC and LHC energies.

\end{abstract}
\pacs{25.75.-q, 25.75.Dw}

\maketitle

\section{Introduction}

High energy nuclear collisions at the Large Hadron Collider at CERN and at the Relativistic Heavy Ion Collider at Brookhaven National Laboratory have created a hot and dense matter of deconfined quarks and gluons: the quark-gluon plasma (QGP)~\cite{Arsene:2004fa,Back:2004je,Adams:2005dq,Adcox:2004mh,RevModPhys.89.035001,BRAUNMUNZINGER201676,CHEN20181}. Because the mass of strange quark has a similar magnitude as the QGP phase transition temperature, strange quarks can be abundantly produced in the QGP. This has led to the suggestion that the production of strange hadrons would be enhanced in relativistic heavy ion collisions~\cite{Rafelski:1982pu,Koch:1986ud}. However, the abundances of various strange hadrons relative to that of pions measured in heavy-ion collisions from SPS to RHIC and LHC energies do not show a significant dependence on either the collision centrality or the collision energy~\cite{Abelev:2007xp}, except for the more pronounced production of multistrange baryons in these collisions. In particular, the ALICE Collaboration has recently observed a significant strangeness enhancement in high multiplicity $pp$ collisions at $\sqrt{s}$ = 7 TeV~\cite{ALICE:2017jyt}. These data also indicate that the multistrange hadron yields relative to that of pions increase significantly with the multiplicity of the collision and the strangeness content of the hyperon as found in $p$-Pb collisions at the same multiplicity densities~\cite{ALICE:2017jyt}. These results for small systems are qualitatively consistent with the predictions of the statistical model~\cite{Redlich2002} that takes into account the effect of canonical suppression due to the conservation of strangeness~\cite{Hagedorn:1984uy,Ko:2000vp} and of the core-corona model that assumes thermal strange production in the core~\cite{Becattini:2008yn,PhysRevC.79.064907}. However, the commonly used Monte-Carlo (MC) models for $pp$ collisions~\cite{Sjostrand:2007gs,PhysRevC.92.034906,PhysRevD.92.094010} have failed to describe these data satisfactorily~\cite{ALICE:2017jyt}. Therefore, the development of a comprehensive microscopic model for understanding the enhanced strangeness production in collisions of small systems is needed.

In this paper, we extend the string melting version of a multiphase transport (AMPT) model by introducing additional coalescence factors for the production of strange baryons in order to describe their enhancements in relativistic heavy-ion collisions from RHIC to LHC energies. This extended AMPT model is found to also describe reasonably well the multiplicity dependence of strangeness enhancement observed in $p$-Pb collisions at the LHC energy.
 
\section{The AMPT model}

The AMPT model~\cite{PhysRevC.72.064901} is a multiphase transport model that has been extensively used for studying relativistic heavy-ion collisions. In this model, the initial conditions are taken from the spatial and momentum distributions of minijet partons and soft string excitations from the ${\rm HIJING}$ event generator~\cite{PhysRevD.44.3501}, which is followed by two-body elastic parton scatterings using the parton cascade model ${\rm ZPC}$~\cite{ZHANG1998193}, the conversion of partons to hadrons via either the string fragmentation or a quark coalescence model, and hadronic scatterings based on the hadronic transport model ${\rm ART}$~\cite{PhysRevC.52.2037}. The default version of the AMPT model~\cite{Zhang:1999bd}, which involves only minijet partons from HIJING in the parton cascade and uses the Lund string fragmentation~\cite{Sjostrand:2007gs} for hadronization, can well describe the rapidity distributions and transverse momentum spectra of identified particles observed in heavy-ion collisions at both SPS and RHIC. It, however, significantly underestimates the elliptic flow at RHIC~\cite{Lin:2001zk}. By converting the initial hadrons produced from the Lund string fragmentation in the default model to their valence quarks and using a simple quark coalescence model to covert them back to hadrons after the ZPC, the string melting version of the AMPT model~\cite{Lin:2001zk} was able to describe the anisotropic flows in both large and small collision systems~\cite{Lin:2001zk,PhysRevLett.113.252301,MA2014209}, although it still failed to describe the proton rapidity distribution and the $\rm p_T$ spectra of hadrons including the pion~\cite{PhysRevC.72.064901}. By further tuning the Lund string fragmentation parameters in the AMPT model, the pion and kaon rapidity distribution, $\rm p_T$ spectra, and elliptic flow at RHIC and LHC energies could also be reasonably described~\cite{PhysRevC.90.014904}. With the refinement of the quark coalescence algorithm for hadronization, the AMPT model has further improved its description of the experimental data, especially the baryon $\rm p_T$ spectra and the antibaryon over baryon ratios~\cite{He:2017tla}. On the other hand, all versions of the AMPT model, including the recent AMPT model with the new quark coalescence algorithm (v1.31t1/2.31t1)~\cite{He:2017tla}, have problems in reproducing the yields and $\rm p_T$ spectra of multistrange baryons in heavy ion collisions~\cite{Pal:2001zw,He:2017tla,Ye:2017ooc,Jin:2018lbk}. For example, the $\rm \Omega^-$ yield from the default AMPT model~\cite{Pal:2001zw} is a factor of two smaller than the Pb+Pb data measured at the SPS energy of 158 AGeV.

\section{New coalescence factors for multistrange baryons production in AMPT}

In general, there are two sources for the production of hyperons in the AMPT model: the initial ones produced directly from quark coalescence after the ZPC in the string melting version (or from the Lund string fragmentation in the default version), and the secondary ones from the strangeness-exchange reactions between baryons and strange mesons in ART~\cite{Pal:2001zw}. Recently, a new version of the AMPT model (v1.31t1/2.31t1) has been developed in Ref.~\cite{He:2017tla} by improving the quark coalescence algorithm in the model. In this new quark coalescence model, a quark searches all antiquarks and records the closest one in relative distance ($\rm d_M$) as the potential coalescence partner to form a meson. It also searches all other quarks and records the two having the closest average distance ($\rm d_B$) as the potential coalescence partners to form a baryon. Given both its meson partner and baryon partners, a quark can then form a meson or a baryon according to the following criterion:
\begin{eqnarray}
\rm d_B < d_M * r_{BM} &:& \rm form~a~baryon; ~~~\rm otherwise~form~a~meson,
\label{eq1}
\end{eqnarray}
where $\rm r_{BM}$ is the baryon coalescence parameter~\cite{He:2017tla} that controls the probability for a quark or an antiquark to form a baryon or an antibaryon relative to that to form a meson. We note that in the string melting version of the AMPT model quarks are converted to hadrons by coalescence after they freeze out kinetically or have had their last scatterings, with the species of the formed hadron being determined by the flavors of coalescing quarks and their invariant mass. In contrast to the original quark coalescence in the AMPT model that forced the numbers of mesons, baryons, and antibaryons in an event to be separately conserved by the quark coalescence, this method only requires the conservation of the net-baryon number in each event.

To better describe the hyperon yields with the new AMPT model of Ref.~\cite{He:2017tla}, we introduce an additional $\rm r_{Y}$ factor to  Eq.(\ref{eq1}) when it is possible for a quark to form a hyperon $Y$ ($\Lambda$, $\Xi$, and $\Omega$) during the coalescence process via the following criterion:
\begin{eqnarray}
\rm d_B < d_M * r_{BM} * r_{\rm Y} : form~a~hyperon; ~~~\rm otherwise~form~a~meson.
\end{eqnarray}
Note that the above criterion requires at least one strange quark or antiquark among the three quarks or antiquarks for the possible baryon or antibaryon formation. For all other cases, Eq.(\ref{eq1}) is used to determine the formation of non-strange baryons.

We also make another change to the string melting AMPT model in order to correct the different initial abundance of strange and antistrange quarks at midrapidity. Despite the zero net-strangeness in the produced partonic matter, strange quarks and antiquarks in string melting AMPT come from the decomposition or melting of strange mesons and hyperons from the Lund string fragmentation, which produces hyperons and antihyperons with different rapidity distributions due to the nonzero net-baryon number. As a result, the yield of antistrange quarks at midrapidity from AMPT is slightly larger than that of strange quarks, and this may result in a larger yield of midrapidity antistrange baryons than strange baryons in heavy-ion collisions at low energies, which contradicts to what is observed in experiments~\cite{Adamczyk:2015lvo}. To resolve this problem, we randomly exchange the initial strange and antistrange quarks produced from the same melting string to make their distributions to be the same. Note that the need to randomize the strange and antistrange quark distributions in the AMPT was discussed earlier in Ref.~\cite{Song:2012cd}. 

\section{Results and Discussions}

Using the AMPT model with these new improvements, we have studied hadron production in Pb+Pb collisions at $\rm \sqrt{s_{NN}}$ = 2.76 TeV and in Au+Au collisions at $\rm \sqrt{s_{NN}}$ = 200 GeV. For the parameters in Lund string fragmentation and the baryon coalescence parameter $\rm r_{BM}$, we adopt the values used in Ref.~\cite{PhysRevC.90.014904} and in Ref.~\cite{He:2017tla}, respectively. We also use the strong coupling constant $\rm \alpha_s$ = 0.33 and a parton cross section of 1.5 mb for the parton cascade ZPC~\cite{He:2017tla}. We then determine the value of the hyperon enhancement factor $\rm r_{Y}$ according to the total yield of hyperon of a given species from experiments~\cite{Abelev:2013xaa,Adam:2015vsf,Adams:2006ke,Adamczyk:2015lvo,Adam:2019koz}. In Table~\ref{Tab1}, the values for ${\rm r}_\Lambda$, ${\rm r}_\Xi$, and ${\rm r}_\Omega$ as well as those of the parameters $a$ and $b$ in Lund string fragmentation are shown for above collision systems as well as for p-$Pb$ collisions at the LHC. 

\begin{table*}[htbp]
\centering

\caption{Values of $a$ and $b$ parameters in the Lund string fragmentation as well as the $\rm r_{BM}$ parameter for baryon production and the $\rm r_{Y}$ factors for hyperon production in the coalescence model for central (0-5\%) Au+Au collisions at the top RHIC energy and central (0-5\%) Pb+Pb collisions or $p$-Pb collisions at the LHC energies.}
\begin{tabular}{cccccccc}
\hline
{\rm System} ~~~ & $\sqrt{s_{\rm NN}}~ ({\rm GeV})$~~~ & $a$~~~ & $b (\rm GeV^{-2})$~~~&$\rm r_{BM}$~~~      &$r_{\Lambda}$~~~  &$r_{\Xi}$~~~  &$r_{\Omega}$ \\
\hline  \hline
{\rm Au+Au} ~~~ & 200                                  & 0.55 & 0.15                  &0.61~~~                  &1.1~~~        &1.2~~~        &1.5    \\
{\rm Pb+Pb} ~~~ & 2760                                & 0.30 & 0.15                &0.61~~~                  &1.1~~~        &1.2~~~        &1.5    \\
{\it p\rm +Pb} ~~~ & 5020                               & 0.30 & 0.15                   &0.54~~~                  &1.2~~~        &1.2~~~        &1.3   \\
\hline
\end{tabular}
\label{Tab1}
\end{table*}

\begin{figure}[!htb]
\centering
\includegraphics[scale=0.40]{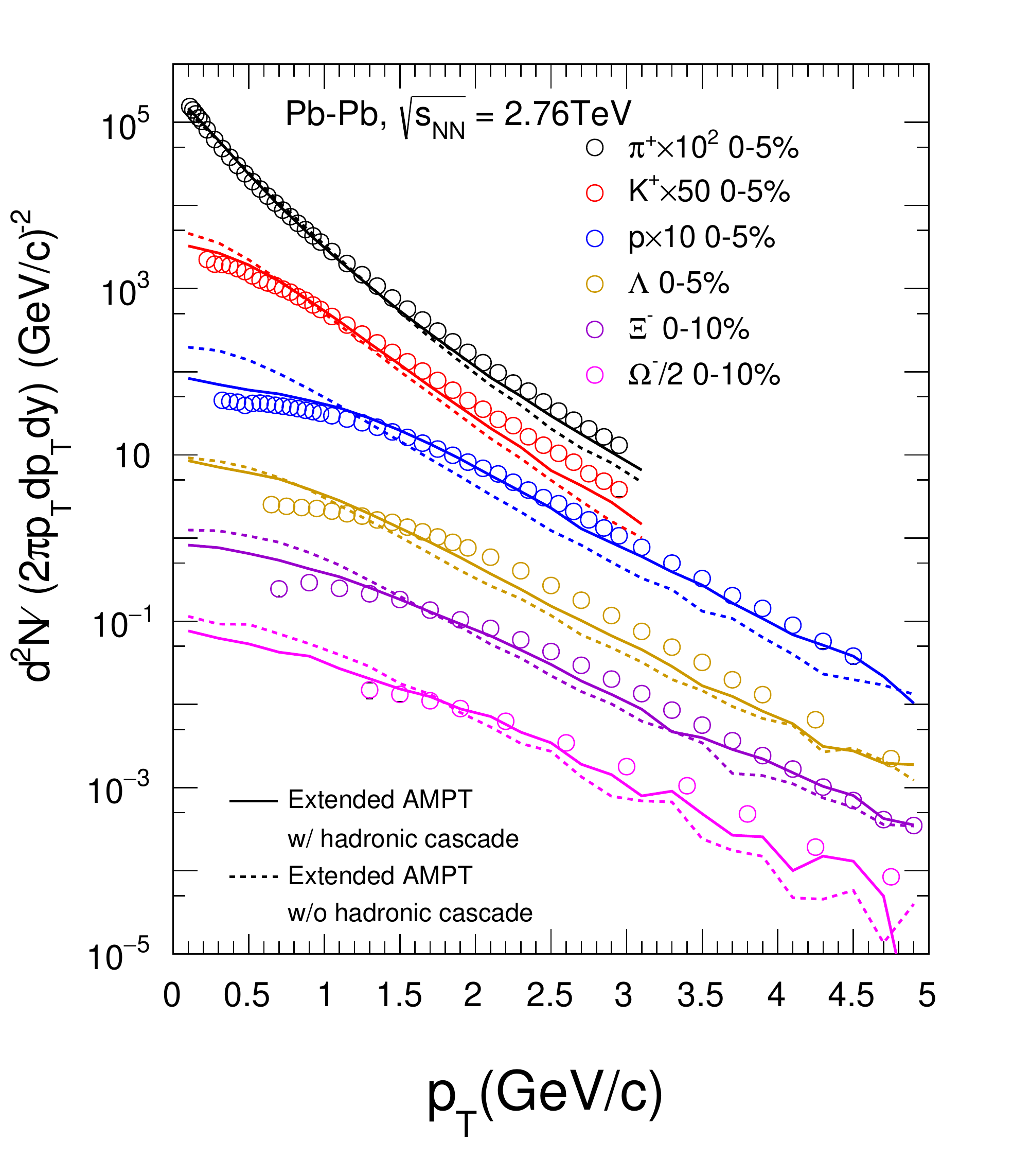}
\includegraphics[scale=0.40]{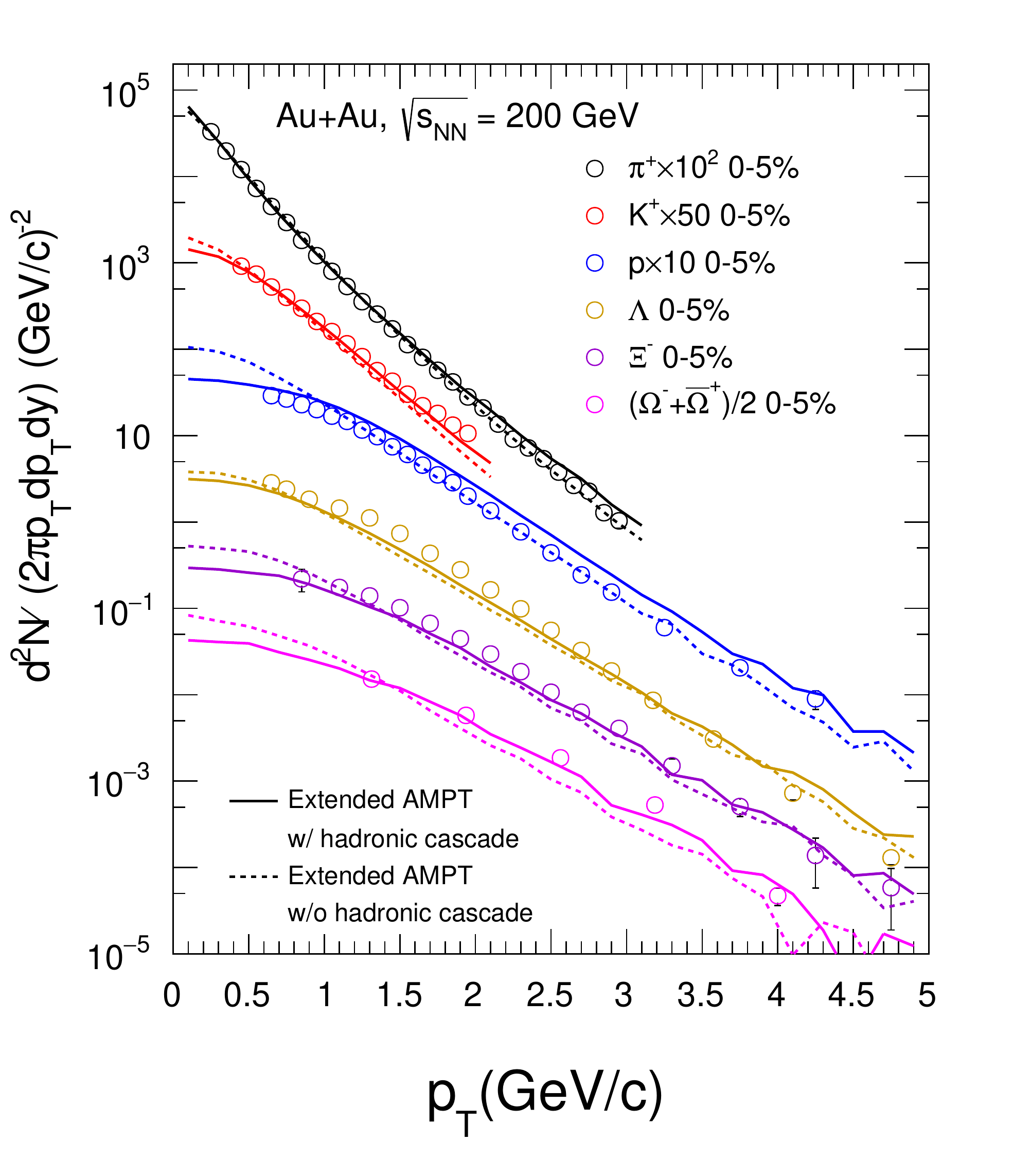}\\
\caption{The midrapidity $\rm p_T$ spectra of $\pi$, K, p, $\Lambda$, $\Xi$ and $\Omega$ in central Pb+Pb collisions at $\rm \sqrt{s_{NN}}$=2.76 TeV (left window) and in central Au+Au collisions at $\rm \sqrt{s_{NN}}$=200 GeV (right window). Open symbols represent the experimental data~\cite{Abelev:2013haa,Abelev:2013xaa,ABELEV:2013zaa,Adam:2015vsf,Adams:2006ke,Adamczyk:2015lvo,Adamczyk:2017iwn,Adam:2019koz}, and the lines are AMPT results with (solid lines) or without (dashed lines) hadron cascade.}
\label{spectra}
\end{figure}

Our results for these two heavy-ion collision systems obtained with the same set of $\rm r_{BM}$ and $\rm r_Y$ values [c.f. Table~\ref{Tab1}] are shown in Fig.~\ref{spectra}. One sees that the extended AMPT model with the $\rm r_Y$ value taken to be 1.1 for $\Lambda$ and larger for $\Xi$ and $\Omega$ not only provides a good description of the measured $\pi$, K, and p transverse momentum $\rm p_T$ spectra like before~\cite{He:2017tla}, it also describes well those of multistrange baryons in heavy ion collisions at both RHIC and LHC energies.

We have also studied the hadronic cascade effect on hadron production at midrapidity in AMPT by turning on and off the hadron cascade component in AMPT, and the results are shown as solid and dashed lines in Fig.~\ref{spectra}, respectively. It is seen that the hadronic effect is not small, resulting in a reduction of $\Xi$ and $\Omega$ yields up to 20\% and the proton yield by 10\%, while an increase of the $\Lambda$ yield by 10\%. Note that these hadronic effects on hyperons depend sensitively on the input hadronic cross sections in the transport model and further detailed studies are needed.

\begin{figure}[ht]
\centering
\includegraphics[scale=0.5]{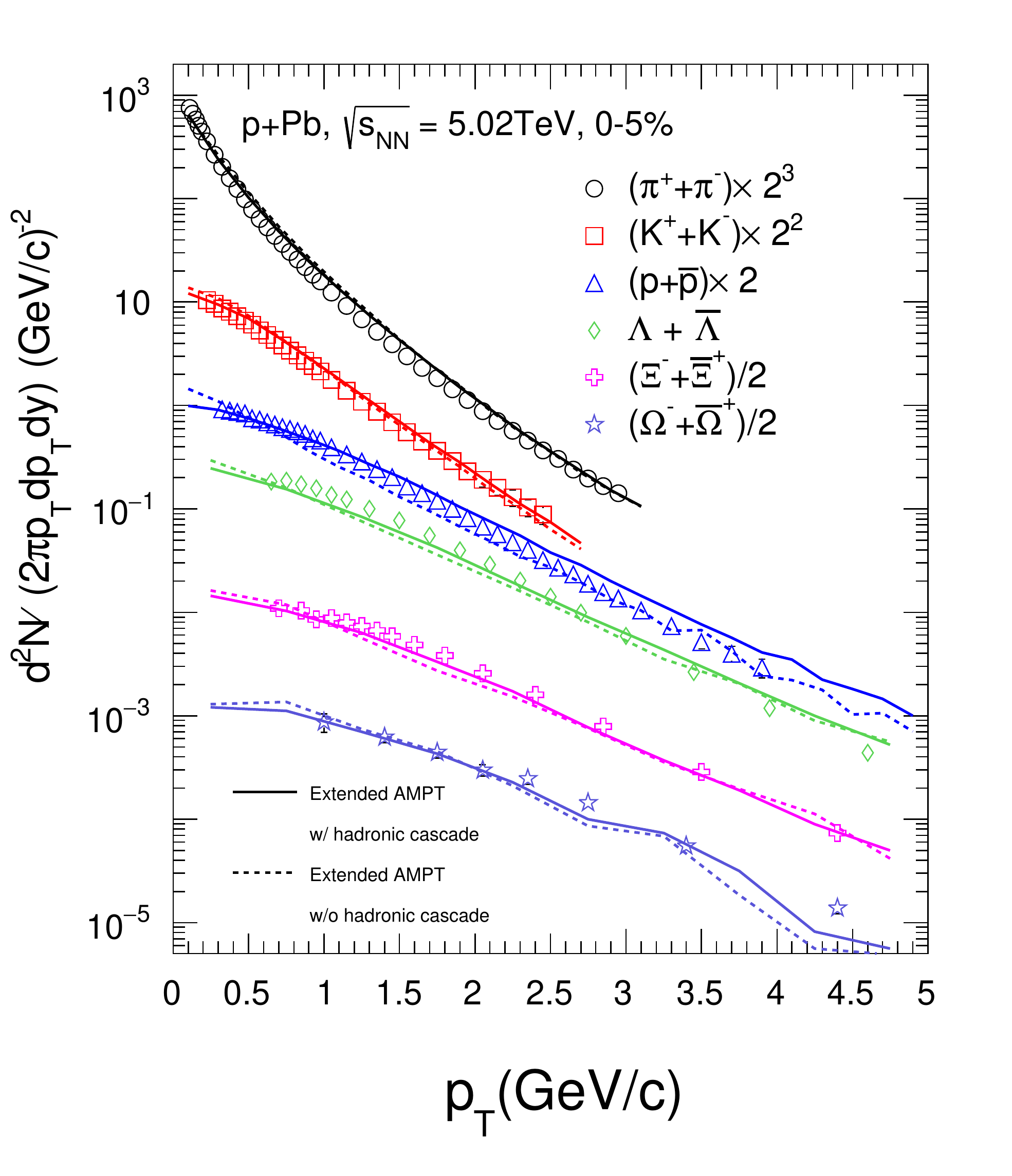}
\caption{The midrapidity hadron $\rm p_T$ spectra in high multiplicity $p$-Pb collisions at $\rm \sqrt{s_{NN}}$= 5.02 TeV. Symbols represent the experimental data for $\pi$, K, p and $\Lambda$ at $\rm 0<y_{CM}<0.5$~\cite{Abelev:2013haa} and for $\Xi$ and $\Omega$ at $\rm -0.5<y_{CM}<0$ ~\cite{Adam:2015vsf}, and the lines are corresponding results from the AMPT model with (solid) or without (dashed) hadron cascade.}
\label{pPb}
\end{figure}

For $p$-Pb collisions at $\rm \sqrt{s_{NN}}$ = 5.02 TeV, we find that a smaller value for the baryon coalescence factor of $\rm r_{BM}$ = 0.54 is required to describe the proton yield~\cite{Abelev:2013haa} because the baryon to meson ratio in high multiplicity $p$-Pb collisions is smaller than that measured in central Pb+Pb collisions~\cite{Abelev:2013haa}. Using the hyperon enhancement factors of $\rm r_{\Lambda}$ = 1.2, $\rm r_{\Xi}$ = 1.2 and $\rm r_{\Omega}$ = 1.3 as given in Table~\ref{Tab1}, the AMPT results for the $\rm p_T$ spectra of $\pi$, K, p, $\Lambda$, $\Xi$ and $\Omega$ in high multiplicity $p$-Pb collisions at $\rm \sqrt{s_{NN}}$ = 5.02 TeV are shown in Fig.~\ref{pPb}, which are seen to agree well with the experimental data. The hadronic cascade effect is seen to be small from comparing the results with and without the hadron cascade component in AMPT, shown as solid lines and dashed lines in Fig.~\ref{pPb}, respectively. We have also checked that the results obtained using the same $\rm r_{Y}$ factors for other multiplicity intervals also describe the hyperon data reasonably well with less than $10$\% difference in the integrated yield at midrapidity. 

\begin{figure}[!htb]
\centering
\includegraphics[scale=0.40]{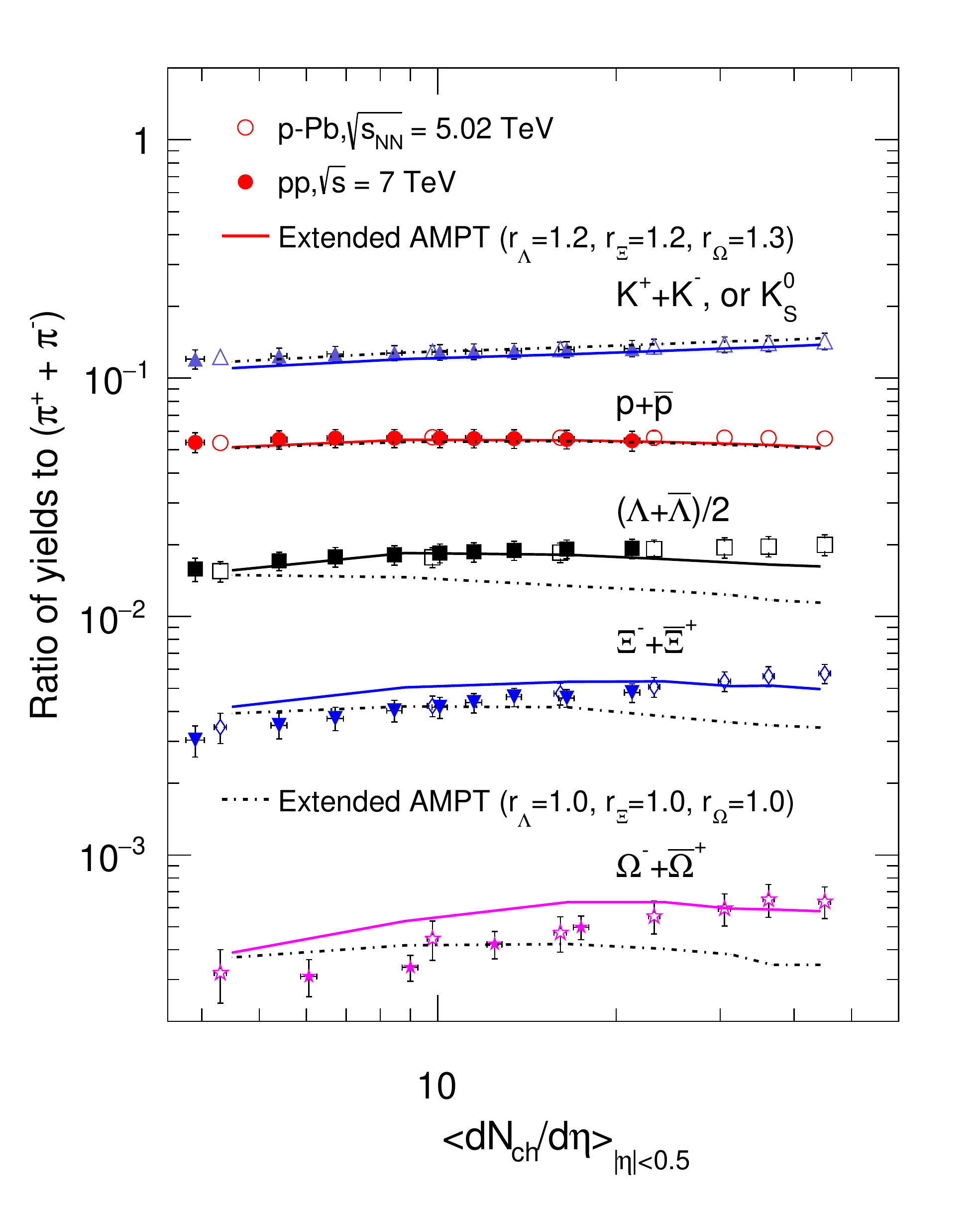}
\caption{The midrapidity yield ratios of kaon, proton, $\Lambda$, $\Xi$, $\Omega$ to pion in $pp$ collisions at $\rm \sqrt{s_{NN}}$ = 7 TeV and in $p$-Pb collisions at $\rm \sqrt{s_{NN}}$ = 5.02 TeV. Results from experiments~\cite{ALICE:2017jyt,Abelev:2013haa,Adam:2015vsf} are shown by open symbols for $p$+Pb collisions and filled symbols for $pp$ collisions, and the lines are AMPT results for $p$-Pb collisions with (solid) or without (dot-dashed) the extra hyperon enhancement $\rm r_Y$ factors.}
\label{ratios}
\end{figure}

Figure~\ref{ratios} presents the multiplicity dependence of $K$, $p$, $\Lambda$, $\Xi$ and~$\Omega$ to pion $\rm dN/dy$ ratios in $p$-Pb collisions at $\rm \sqrt{s_{NN}}$ = 5.02 TeV based on two million events from the extended AMPT model for each multiplicity interval. For the criteria on multiplicity interval selection, we follow the experimental method by using the charged particle pseudorapidity distributions within $|\eta| <$ 0.5. Our $\langle dN_{ch}/d\eta \rangle$ values for the centrality of 0-5\%, 5-10\%, 10-20\%, 20-40\%, 40-60\%, 60-80\%, and 80-100\% are 46, 37, 32, 24, 16, 8, and 3, respectively, which are consistent with the values used in experiments~\cite{Abelev:2013haa,ALICE:2017jyt}. The AMPT results obtained with the $\rm r_{Y}$ factors as given in Table~\ref{Tab1} (solid lines) are seen to describe very well the $K$ and $p$ to pion yield ratios and reasonably well the hyperon to pion yield ratios. On the other hand, without the hyperon enhancement  $\rm r_{Y}$ factors (i.e., using $\rm r_{Y}$ = 1 for all hyperons) leads to a significant underestimation of the measured hyperon yields as shown by the dash-dotted lines, especially for more central $p$-Pb collisions. A comparison of the two sets of theoretical results shows that the enhancement in the hyperon yields due to the $\rm r_Y$ factors depends both on the multiplicity in a collision and the strangeness content of the hyperon. For $\Lambda$, the enhancement is up to 50\% in central and mid-central $p$-Pb collisions and 30\% at 80-100\% centrality. For $\Xi$ that has the same $\rm r_{Y}$ as for $\Lambda$ according to Table~\ref{Tab1}, these numbers are 40\% for central and mid-central collisions, 30\% at 60-80\% centrality, and 20\% at 80-100\% centrality. The enhancement in $\Omega$ production due to $\rm  r_\Omega$ shows a similar centrality dependence as the $\Xi$. We note that the hadronic effect in $p$-Pb collisions is generally important for proton due to baryon-antibaryon production from and annihilation to mesons, and the yields of various hyperons are only slightly affected by the strangeness-exchange reactions in the hadronic matter. It is worth mentioning that including the hadronic effect is important in describing the multiplicity dependence of short-lived resonances, such as the $\rm K^{*0}$ meson yield measured in $p$-Pb collisions at LHC energy~\cite{Liu_2017}.

\begin{figure}[ht]
\centering
\includegraphics[scale=0.4]{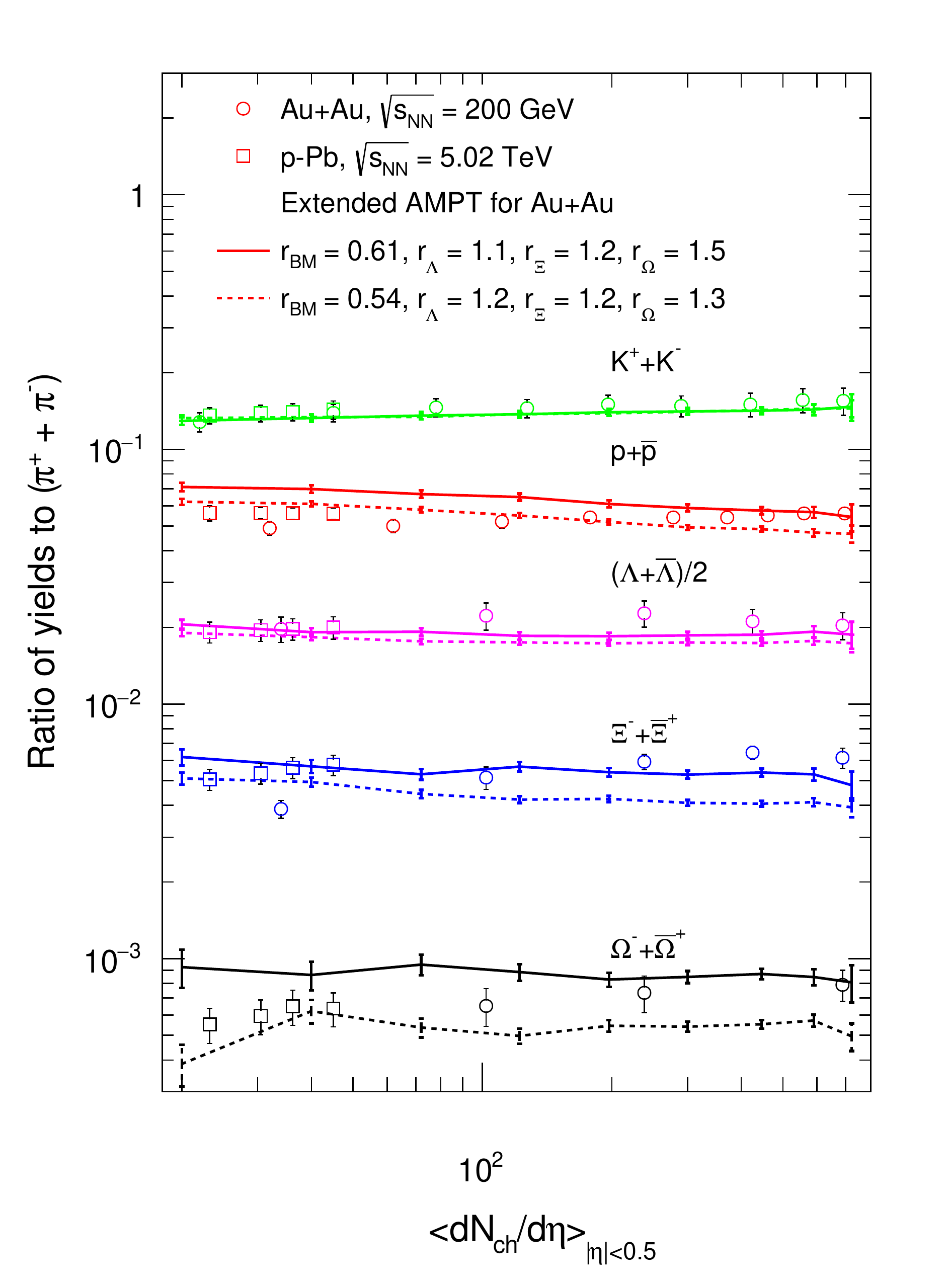}
\caption{The multiplicity dependence of $K$, $p$, $\Lambda$, $\Xi$ and $\Omega$ to pion yield ratios in Au+Au collisions at $\rm \sqrt{s_{NN}}$ = 200 GeV from the extended AMPT model using different values of coalescence factors (solid and dashed lines). Results from experiments are shown by open circles for Au+Au collisions~\cite{Adams:2006ke,Adler:2003cb} and open 
squares for p-Pb collisions~\cite{Abelev:2013haa,Adam:2015vsf}.}
\label{AA-ratios}
\end{figure}

The need to use different values of the coalescence factors ($\rm r_{BM}$ and $\rm r_Y$) for $p$-Pb collisions and central Pb+Pb collisions at the LHC energies suggests that these phenomenological parameters depend on the system size. Such a phenomenon is also present in a recent study~\cite{PhysRevC.101.024912}, in which a multiplicity dependent parameter is introduced in a dynamical core-corona model to describe the measured strange baryons to pion ratios in relativistic heavy-ion collisions. This is not surprising since the coalescence model used in the present AMPT model for hadronization is based on formation of hadrons from nearest neighbor quarks, instead of using the hadron Wigner functions as in more realistic approach~\cite{Chen:2006vc,Wang:2019xph}. Also, in ordering the sequence for partons to coalesce in the AMPT, it is currently based on the inverse of the freeze-out time, using other types of ordering could affect the values of $\rm r_{BM}$ and $\rm r_Y$.

The dependence of the coalescence factors on system size should also affect the centrality dependence in A+A collisions, therefore we have further studied the centrality dependence of $K$, $p$, $\Lambda$, $\Xi$ and $\Omega$ to pion yield ratios in Au+Au collisions at $\rm \sqrt{s_{NN}}$ = 200 GeV. Shown in Fig.~\ref{AA-ratios} by solid and dashed lines are results from the extended AMPT model by taking the values of the four coalescence factors from Table~\ref{Tab1} for A+A collisions and $p$-Pb collisions, respectively; while the values of the parameters $a$ and $b$ in the Lund string fragmentation are taken from Table~\ref{Tab1} for Au+Au collisions. 
For comparison, we show by open symbols the experimental data from Au+Au collisions at $\rm \sqrt{s_{NN}}$ = 200 GeV~\cite{Adams:2006ke,Adler:2003cb} and the high-multiplicity events of $p$-Pb collisions at $\rm \sqrt{s_{NN}}$ = 5.02 TeV~\cite{Abelev:2013haa,Adam:2015vsf}. We see that the extended AMPT using the coalescence parameter values for A+A collisions in Table~\ref{Tab1} (solid lines) gives a good description of the data from central Au+Au collisions but systematically over-predicts the $p$, $\Xi$ and $\Omega$ data from peripheral collisions. On the other hand, the extended AMPT using the coalescence parameter values for $p$-Pb collisions (dashed lines) improves its description of the experimental data for peripheral Au+Au collisions. 
The observation that peripheral and central Au+Au collisions prefer different sets of $\rm r_{BM}$ and $\rm r_Y$ values shows that these coalescence factors depend on the size of a collision system. 
Furthermore, the improvement in the description of peripheral Au+Au collisions using the coalescence parameter values for $p$-Pb collisions 
suggests that the coalescence factors do not depend much on whether the system is produced from A+A or p+A collisions. 
Note that the possible dependence of the coalescence factors on the collision energy (at the same system size) is not investigated in this study.

\begin{figure}[!htb]
\centering
\includegraphics[scale=0.4]{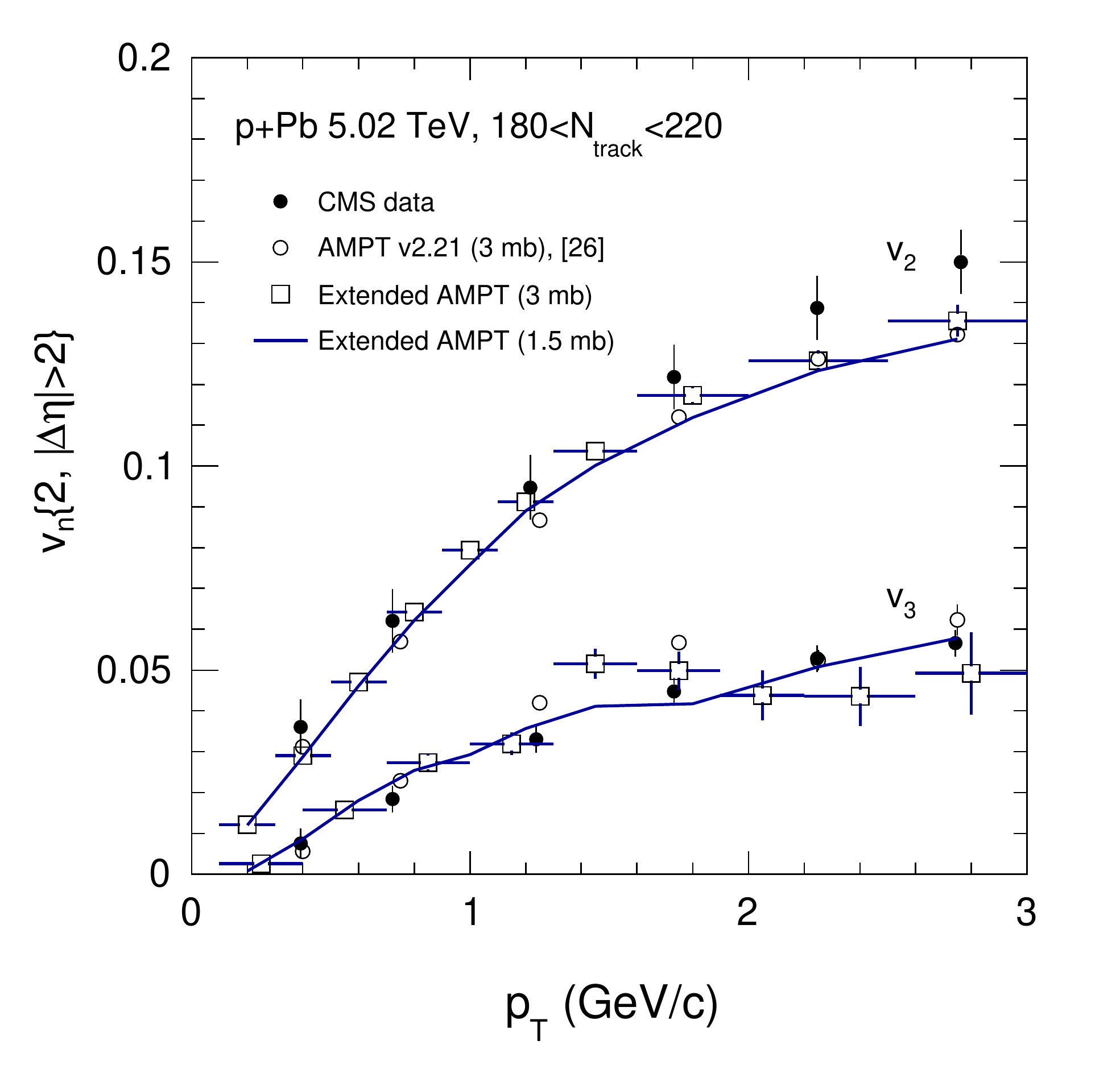}
\caption{The $\rm p_T$ dependence of elliptic flow $v_2$ and triangular flow $v_3$ of charged particles in high multiplicity $p$-Pb collisions from the extended AMPT model with parton cross sections of 1.5 mb (solid lines) and 3 mb (open squares), respectively. The CMS data are represented by filled circles, and results from an early version of AMPT model are denoted by open circles.}
\label{flow}
\end{figure}

As in an early version of the AMPT model~\cite{PhysRevLett.113.252301,MA2014209} that used the original quark coalescence algorithm, the extended AMPT model in the present study can also describe the differential elliptic flow $\rm v_2(p_T)$ and triangular flow $\rm v_{3}(p_T)$ as functions of $\rm p_T$. Results obtained from the long-range two-particle azimuthal correlation functions are shown in Fig.~\ref{flow} for charged particles with $|\eta| < $ 2.4 and $\rm p_T >$ 0.1 GeV/c in a high multiplicity window. Our results with parton scattering cross section of 1.5 mb and 3 mb, shown by solid lines and open squares, respectively, are close to each other, and both give a good description of the experiment data shown by filled circles~\cite{Chatrchyan:2013nka}. Compared to the earlier AMPT results~\cite{PhysRevLett.113.252301} (open circles), results from present AMPT calculations using the same parton scattering cross section are about the same. The seemingly non-monotonic behavior of $\rm v_3(p_T)$ from the parton scattering cross section of 3 mb in our results is largely due to statistical fluctuations as the number of events in our study is much smaller than that in Ref.~\cite{PhysRevLett.113.252301}.
However, the present extended AMPT model describes better the baryon yields and $\rm p_T$-spectra, while the early results~\cite{MA2014209} under-predicted the low-$\rm p_T$ yield of pions but over-predicted that of protons. Although the hadron $\rm p_T$ spectra shown in Figs.~\ref{spectra} and \ref{pPb} are obtained from the extended AMPT model with a parton scattering cross section of 1.5 mb, using a larger cross section of 3 mb only has a very small effect on these results.

\section{Conclusion}

We have introduced additional coalescence factors for strange baryons in the string melting version of the AMPT model in order to improve its description of multistrange hadron production in heavy ion collisions at both the LHC and RHIC energies. This extended AMPT model is shown to describe reasonably well the hyperon $\rm p_T$ spectra in central Pb+Pb collisions at the LHC energy and Au+Au collisions at the top RHIC energy. It can also qualitatively describe the multiplicity dependence of strangeness enhancement in high multiplicity $pp$ and $p$-Pb collisions at LHC energies. We find that the coalescence factors depend on the system size but not much on where the system is produced from, and using multiplicity-dependent coalescence factors improves the description of the centrality dependence of particle production in Au+Au collisions. The current study provides a convenient way to model the mechanism underlying the strangeness enhancement observed in both small and large systems from nuclear collisions at the LHC, and it could also be relevant for the study of the QCD phase diagram in the Beam Energy Scan Program at RHIC via strange baryon production~\cite{Shao:2019xpj}.

\section{acknowledgements}

The work of T.S. and J.C. was supported in part by the Strategic Priority Research Program of Chinese Academy of Sciences with Grant No. XDB34030200 and the National Natural Science Foundation of China under Contract Nos. 11890710, 11775288, 11421505 and 11520101004, while that of C.M.K. was supported in part by the US Department of Energy under Contract No. DE-SC0015266 and the Welch Foundation under Grant No. A-1358.

\bibliography{myref_v0}

\end{document}